\documentclass[a4paper,12pt]{article}
\usepackage[
    left=2cm,
    right=2cm,
    top=3cm,
    bottom=3cm
]{geometry}
\usepackage{natbib}
\usepackage{hyperref}
\hypersetup{
    colorlinks = true,
    urlcolor   = blue,
    citecolor  = black,
}
\usepackage[utf8]{inputenc}
\usepackage{graphicx}
\usepackage{amssymb}
\usepackage{amsmath}
\usepackage{bm}      
\usepackage{booktabs}   
\usepackage{url}        
\usepackage{overpic}   
\usepackage{array}     
\newcolumntype{P}[1]{>{\centering\arraybackslash}p{#1}}
\usepackage{tabularx}
\usepackage{titlesec}
\usepackage{xcolor}

\graphicspath{{./figure/}}

\newcommand{\media}[1]{{\langle #1 \rangle}}
\newcommand{\de}{{\rm{d}}}

\newcommand{\Pra}{{\text{Pr}}}

\newcommand{\Fr}{{\text{Fr}}}
\newcommand{\Ri}{{\text{Ri}}}

\newcommand{\mediap}[1]{{\langle #1 \rangle}_p}
\newcommand{\Order}{{\cal O}}

\title{Thermal behaviour of settling heavy inertial particles in turbulent natural convection}
\author{Hamid Reza Zandi Pour, Guido Boffetta, Stefano Musacchio \and Filippo De Lillo\footnote{Corresponding author: \href{mailto:filippo.delillo@unito.it}{filippo.delillo@unito.it}}
\\[1ex]
  {\small
  Department of Physics and INFN, University of Torino, via P. Giuria 1, 10125 Torino, Italy
  }
}
\begin{document}
\maketitle

\abstract{We investigate the dynamical and thermal behaviour of settling heavy
inertial particles in bulk turbulent thermal convection. Direct numerical
simulations are performed in a homogeneous Rayleigh–Bénard convection setup, where a uniform mean temperature gradient drives statistically
homogeneous convective turbulence seeded with passive inertial particles.
We quantify the effects of dynamic and thermal inertia
on particle statistics. In particular, we investigate how the fluctuations of particle velocity and temperature can be traced back to the correlations between particle and fluid variables, and how particle temperature statistics is affected by the interplay of settling and preferential concentration.}

\section{Introduction}

The interplay between heavy inertial particles and turbulent thermal convection is
central to many natural and industrial phenomena, from cloud formation
\citep{grabowski2013growth,kumar2013cloud,kumar2014lagrangian} and pollutant
dispersion \citep{britter2003flow} to combustion \citep{battista2011intermittent}.  
Inertial particles denser than the advecting fluid tend to cluster in the high-strain regions separating vortical structures \citep{eaton1994preferential,bec2007heavy}. This preferential concentration correlates with temperature fluctuations, since a scalar field advected by turbulence tends to form in the same high-strain regions \citep{bec2014clustering}. 
When the effects of gravity are considered, turbulence affects the settling velocity of particles, generally increasing it as a result of a preferential sampling of downward fluid velocities
\citep{bec2014gravity, Ireland2016b, maxey1987gravitational, wang1993settling,
ayala2008effects,delillo2008sedimentation,rosa2016settling}. This increase has a maximum at intermediate inertia, while very heavy particles should settle with the still-fluid velocity due to asymptotic decorrelation from the flow. Small reduction effects are observed in some simulations and experiments when the settling speed is of the order of the typical velocity fluctuation \citep{clementi2026enhancement}.  
Settling itself, on the other hand, alters turbulence-induced clustering \citep{Bragg2021,bec2014gravity} and therefore the sampling of the temperature field by the particles \citep{li2024temperature}. 

The thermal statistics
of inertial particles has been considered, with~\citep{Carbone2019,Saito2022,Zandipour2022fluids} and without~\citep{li2024temperature, bec2014clustering} feedback on fluid temperature, with the thermal coupling leading to a dampening of the temperature fluctuations in the fluid. 
In many numerical studies, temperature is modelled as a passive scalar field stirred by the turbulent flow~\citep{bec2014clustering}. This model is relevant for the regime of forced convection, in which turbulence is mainly produced by external mechanical forcing while the buoyant force can be neglected.  In this case, there is no correlation between the statistics of the fluid velocity and temperature sampled by the particles. 
Conversely, the regime of natural convection is characterized by a strong correlation between the vertical component of fluid velocity and temperature, since this correlation is responsible for heat transport in the bulk and acts as a source term for both velocity and temperature fluctuations. It is therefore interesting to investigate how the correlation induced by the buoyant force interacts with the 
processes of preferential concentration and settling in determining the statistics of velocity and temperature of heavy particles.
Recent works have examined thermal interactions between fluids and settling heavy inertial particles in forced
isotropic and homogeneous turbulence, showing that settling alters Lagrangian correlations
and influences the emergence of thermal caustics, a common feature of particles with finite
thermal inertia  \citep{ZandipourPOF2024,li2024temperature}.

In this study, we aim at exploring the combined influence of buoyancy, gravity and turbulence on the velocity and temperature statistics of heavy inertial particles settling in a
homogeneous Rayleigh–B\'enard convection (HRBC) \citep{Calzavarini2005,Calzavarini2006}. In this idealized configuration, turbulence is forced by buoyancy due to temperature differences, while thermal energy is injected by means of an unstable temperature gradient applied through the bulk of the periodic numerical volume. This allows us to isolate the essential physics of buoyancy-driven bulk turbulence while
avoiding wall and boundary layer effects. 

We perform direct numerical simulations of thermally and dynamically one-way coupled particles in HRBC, systematically varying the particles' dynamic and thermal inertia, while keeping all fluid parameters constant. Our goal is to uncover how the interplay between settling, inertia, and buoyancy shape particle velocity and temperature statistics. 

The paper is structured as follows: in section~\ref{sec2} the physical model is introduced, together with a few fundamental relations; in section~\ref{sec3} the results are discussed, preceded by an intruduction of the numerical method and a characterization of the Eulerian properties of the flow. The last section summarizes the results.

\section{Physical model}
\label{sec2}

We consider a buoyancy driven turbulent flow governed by Navier-Stokes equations within the Boussinesq approximation. 
The variation of fluid density field $\rho(\bm{x},t)$ is proportional to the fluid temperature field $T(t,\bm{x})$ through its thermal expansion coefficient $\alpha_T$ as $(\rho-\rho_0)/ \rho_0 = -  \alpha_T (T-T_0)$, where 
$\rho_0$ and $T_0$ are a reference density and temperature respectively.

A mean temperature gradient parallel to the direction of gravity, $x_3$, is imposed to mimic unstable thermal stratification and to inject energy into the flow.
The imposed mean temperature gradient $\gamma=\Delta T/L>0$ represents the temperature jump $\Delta T=T_0-T_1$ across the domain of size $L$.
By introducing the temperature fluctuations $\theta(\bm{x},t)$, such that $T(\bm{x},t)=T_0-\gamma x_3+\theta(\bm{x},t)$, 
the Boussinesq equations for the incompressible flow read
\begin{align}
\frac{\partial u_i}{\partial x_i}&=0,
\label{eq1}\\
\frac{\partial u_i}{\partial t}+u_j \frac{\partial u_i}{\partial x_j} &=
-\frac{1}{\rho_0} \frac{\partial p}{\partial x_i}
+\nu \frac{\partial^2 u_i}{\partial x_j x_j}
+\alpha_T \theta g \delta_{i3},
\label{eq2}
\\
\frac{\partial \theta}{\partial t}+u_j \frac{\partial \theta}{\partial x_j}&=
\kappa \frac{\partial^2 \theta}{\partial x_j x_j}
+
\gamma u_3,
\label{eq3}
\end{align}
where $\bm{u}(\bm{x},t)$ is the velocity field, $\nu$ and $\kappa$ are the kinematic viscosity and the thermal diffusivity respectively, while $g$ denotes the gravitational acceleration.

Inertial particles are modeled as spherical point-like mass, with a density much larger than the fluid and with finite thermal inertia under one-way coupling regime, i.e. with no feedback of the particles on the fluid's momentum or temperature.   
A simplified form of the Maxey–Riley–Gatignol equation \citep{Maxey1983, Gatignol1983} is used to derive the particle Lagrangian motion, and additional corrections, e.g., Faxén correction, and finite-size effects are neglected.

Under these conditions, heat exchange between the particle and fluid can be described by a quasi-steady (diffusive) convective model based on Newton’s law of cooling. In this study, heat exchange between the particle and the surrounding fluid occurs exclusively through conduction. Since the solid particle model assumes a constant particle mass, latent heat effects are not included. However, in other situations, additional heat transfer mechanisms, such as radiation, or heat sources and sinks, such as vaporization enthalpy, can be incorporated.
Furthermore, heat conduction within the particle is neglected because of its small size and the particle temperature is considered uniform.
Within these limits, the
equations for the particle position ${\bf x}_p$, velocity ${\bf v}_p$ and relative temperature $\theta_p$ are 
\begin{align}
\frac{\de x_{p,i}(t)}{\de t}&= v_{p,i}(t),
\label{eq4j}\\
\frac{\de v_{p,i}(t)}{\de t}&=-\frac{1}{\tau_s}\left [ v_{p,i} (t)- u_i(\bm{x}_p,t) \right]- g \delta_{i3}, 
\label{eq5}\\
\frac{\de \theta_p(t)}{\de t}&= -\frac{1}{\tau_\theta} \left [ \theta_p(t)- \theta(\bm{x}_p,t)\right] + \gamma v_{p,3},
\label{eq6}
\end{align}
where $\theta_p=T_p-(T_1-\gamma x_{p,3})$ represents the temperature fluctuation of the particle with respect the local equilibrium value. 
The coefficients $\tau_s$ and $\tau_\theta $ are the particle momentum and thermal relaxation times and are given by
 \begin{align}
\tau_s=\frac{2}{9}\frac{\rho_p}{\rho_0}\frac{R^2}{\nu}, \qquad
\tau_\theta=\frac{1}{3} \frac{\rho_p c_{pp}}{\rho_0 c_{p0}} \frac{R^2}{\kappa},
\label{eq7}
\end{align}
where $R$ is the particle radius, $\rho_p$ the particle density and $c_{pp}$ the heat capacity at constant pressure ($c_{p0}$ the one for the fluid).

The relevant dimensionless numbers of the flow are the Froude number $\Fr$, the Reynolds number ${\text{Re}}$ and the Prandtl number $\Pr$
%and the Richardson number, defined as
\begin{align}
\Fr=\frac{U}{\sqrt{g L}}, \qquad
{\text{Re}}=\frac{U L}{\nu}, \qquad
\Pra=\frac{\nu}{\kappa}, \qquad
\label{eq8}
\end{align}
where $U=(\alpha_T g \Delta T L)^{1/2}$ is a characteristic velocity which together with the characteristic scale $L$ and time $\tau_L=L/U$ is used to make the results dimensionless. 
We remark that in this setup the Richardson number is $\Ri=\alpha_T g \Delta T L/U^2=1$ identically.
The inertial  Stokes number $\mathrm{St}_\eta$ and thermal Stokes number $\mathrm{St}_{\theta}$ are defined as the ratio of the relaxation times to the Kolmogorov time scale $\tau_\eta =(\nu/\varepsilon)^{1/2}$, where $\varepsilon$ is the dissipation rate of the turbulent kinetic energy. 
\begin{align}
\mathrm{St}_\eta =\frac{\tau_s}{\tau_\eta}, \qquad
\mathrm{St}_\theta =\frac{\tau_\theta}{\tau_\eta},
\label{eq9}
\end{align}

Before considering the statistical quantities, let us introduce a few notations. Given two fluid observables $X$ and $Y$ and their particle counterparts  $X_p$ and $Y_p$, $\langle X\rangle$ indicates the spatial average over the fluid domain and over an ensemble of independent flow fields, $\langle X\rangle_p$ a Lagrangian average along inertial particle trajectories, $\sigma_{XY}= \media{XY} - \media{X} \media{Y}$ the covariance. Naturally, $\langle X_p\rangle=\langle X_p\rangle_p$ and $\langle X Y_p\rangle$=$\langle X Y_p\rangle_p$, etc.: averages involving at least one particle observable are always Lagrangian averages.
The budget equation for fluid turbulent kinetic energy $E = 1/2 \langle |{\bm u}|^2 \rangle$, temperature variance $\langle \theta^2 \rangle$ and for the correlation between vertical velocity and temperature fluctuations $\langle u_3 \theta \rangle$
can be written as
\begin{align}
\frac{1}{2} \frac{d \langle |{\bm u}|^2 \rangle}{d t} &= \alpha_T g\langle u_3 \theta \rangle - \nu \langle (\partial_i u_j)^2 \rangle,
\label{eq.fluid.TKE}\\
\frac{1}{2} \frac{d \langle \theta^2 \rangle}{d t} &= \gamma \langle u_3 \theta \rangle - \kappa \langle (\partial_i \theta)^2 \rangle,
\label{eq.fluid.TTE}\\
\frac{d \langle u_3 \theta \rangle}{d t} &= \gamma \langle \theta^2 \rangle + \alpha_T g\langle u_3^2 \rangle + \langle p \partial_3 \theta \rangle 
- (\nu+\kappa) \langle \partial_i \theta \partial_i u_3 \rangle,
\label{eq.fluid.TPE}
\end{align}

These equations show how the correlation $\langle u_3 \theta \rangle$ generates, at large scales, turbulent kinetic energy and temperature variance, which are dissipated at small scale by viscosity and diffusivity.
On the other hand, $\langle \theta^2\rangle $ and $\langle u_3^2 \rangle$ act as source terms for the correlation $\langle u_3 \theta \rangle$, generating a mechanism of positive feedback 
which amplifies the fluctuations. This behaviour is particularly important in HRBC and was indeed observed in  \cite{Calzavarini2005,Calzavarini2006} and has been connected to the existence of {\em elevator modes}, or vertical jet-like structures within the flow, which grow exponentially in time and lead to large fluctuations in the thermal flux. The growth of the fluctuations is stopped by instabilities breaking down the jets. As a consequence of such strong fluctuations, statistical convergence of the results requires very long numerical simulations.

The terminal velocity of a particle in a still fluid, $v_g= \tau_s g $ will be used as a reference velocity. 
In analogy with $v_g$ we introduce a terminal settling temperature: if a particle falls in a still fluid with a fixed mean temperature gradient $\gamma$, its temperature will asymptotically deviate from the background by a quantity $\theta_g= \gamma \tau_\theta v_g=\gamma \tau_\theta  \tau_s g$.

In a generic statistically-stationary flow, the steady state conditions can be derived from \eqref{eq5} and \eqref{eq6}
\begin{align}
\mediap{v_{p,3}} & =  \mediap{u_3} -\tau_sg
\label{eq.part.mean.vel}
\\
\mediap{\theta_p} & =  \mediap{\theta} + \tau_\theta \gamma \mediap{v_{p,3}}
\label{eq.part.mean.temp}
\end{align}

The parameter space of the problem is defined by \eqref{eq8} and \eqref{eq9}. In particular, particle trajectories are determined by the flow parameters \eqref{eq8} and particle inertia $\mathrm{St}_\eta$. 
Because of the one-way coupling between particles and fluid, the statistics of fluid quantities (both dynamic and thermal) seen by the particles is also determined by the same set of parameters. Particle temperature, on the other hand, is affected by all the parameters. 

In this work, all the flow parameters are kept constant while particle inertia and thermal inertia are varied in order to explore their impact on the dynamics. The second order moments (variances and covariances) associated to particle velocity and temperature are given by 
\begin{align}
\sigma_{v_{p,3}}^2 &= \sigma_{u_3 v_{p,3}} 
\label{eq.part.var.vel}\\
\sigma_{\theta_p}^2 &= \sigma_{\theta \theta_p} + \tau_\theta \gamma \sigma_{v_{p,3} \theta_p}
\label{eq.part.var.temp}\\
\sigma_{v_{p,3} \theta_p} &= \frac{1}{\tau_s +\tau_\theta} 
\left [ \tau_\theta \sigma_{u_3 \theta_p}
+ \tau_s \sigma_{v_{p,3} \theta}
+\tau_s \tau_\theta \gamma \sigma_{u_3 v_{p,3}  } \right ]
\label{eq.part.var.covar}
\end{align}
As can be seen from \eqref{eq.part.var.vel}--\eqref{eq.part.var.covar}, the equations for particle variances and covariance contain terms depending on fluid temperature and velocity at particle position (e.g., $\sigma_{u v_p }$, $\sigma_{\theta \theta_p}$, etc.).
Going beyond this point would require equations for the statistics of the fluid observables sampled by particles. Such equations could in principle be written formally, but in practice would require further statistical modeling to produce any useful information. We will instead study the relative statistics obtained numerically by DNS.

\section{Discussion of the numerical results}
\label{sec3}

%======= Eulerian field statistics
\begin{figure}
 \begin{overpic}[scale=0.60,percent]{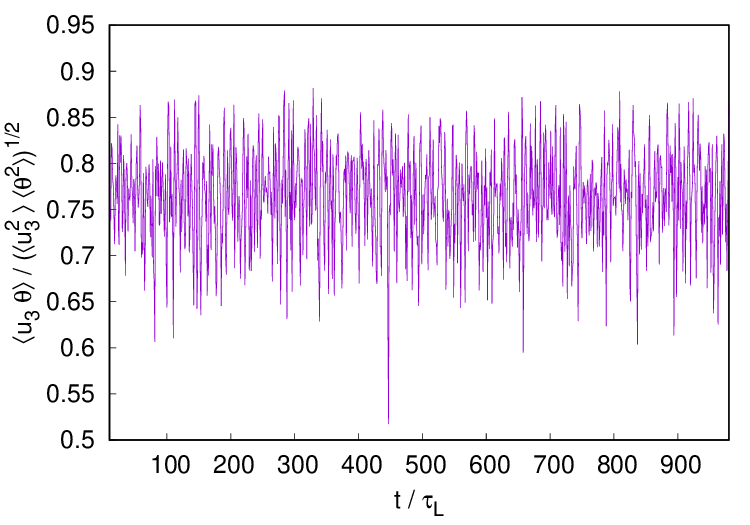}
    \put(85,60){\textbf{(a)}}
  \end{overpic}
    \begin{overpic}[scale=0.60,percent]{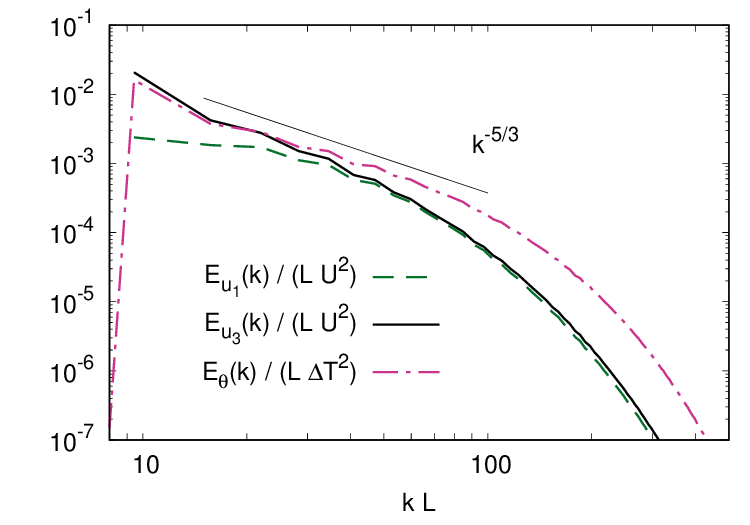}
    \put(85,60){\textbf{(b)}}
  \end{overpic}
 \caption{Characterization of the carrier convective flow. (a) Correlation between vertical velocity component and temperature of the fluid, as a function of time; (b) spectra of horizontal and vertical turbulent kinetic energy and temperature fluctuations.}
 \label{fig:spettri}
\end{figure}

We performed a series of direct numerical simulations of the model by using a pseudo-spectral code in a triply-periodic cubic domain of size $L=2 \pi$ at resolution $256^3$. A standard 2/3 de-aliasing rule is applied with a second-order Runge-Kutta time integration. An imposed mean temperature gradient force the flow at a Reynolds number ${\text{Re}}_{\lambda}=54$.
Viscosity and diffusivity, with $\Pr=1$, are fixed to fully resolve the dissipative scales with $k_{max} \eta \simeq 2.4$ 
and the Froude number is $\Fr=0.1$.
Particles are integrated with the same Runge-Kutta method and fluid fields at particles positions are obtained by tri-linear interpolation scheme.
All simulations are conducted in a statistically stationary state by a long run at least \Order (100) integral time scale $\tau_L$ and Lagrangian statistics are collected after an initial transient time. 

Three sets of simulations are considered, representing three cuts through the parameter space of the dynamic Stokes number $\mathrm{St}_\eta$ and the thermal Stokes number $\mathrm{St}_\theta$, which are varied in the range from 0.1 to 15.
The first case corresponds to particles made of a given material and therefore with the ratio $\mathrm{St}_\eta/\mathrm{St}_\theta$ fixed.
While this is physically relevant, isolating the contributions of particle inertia and thermal inertia is difficult. 
For this reason we also consider two sets of runs in which one Stokes number changes with the other one fixed, enabling a clearer interpretation of their separate effects.
 
In figure~\ref{fig:spettri}a we plot the normalized heat flux which forces the system, 
as a function of time for a typical run while in 
figure~\ref{fig:spettri}b  the spectra of kinetic energy (decomposed in the horizontal and vertical components) and of thermal energy are shown and a narrow inertial range is observable.

%======= Lagrangian particles statistics
\subsection{Lagrangian particles statistics }
\label{sub.sec.Lag.part.stat}
%%%%%%%%%%% new fig vis %%%%%%%%%%
%%% VERTICAL 
\begin{figure}
  \begin{overpic}[scale=0.6,percent]{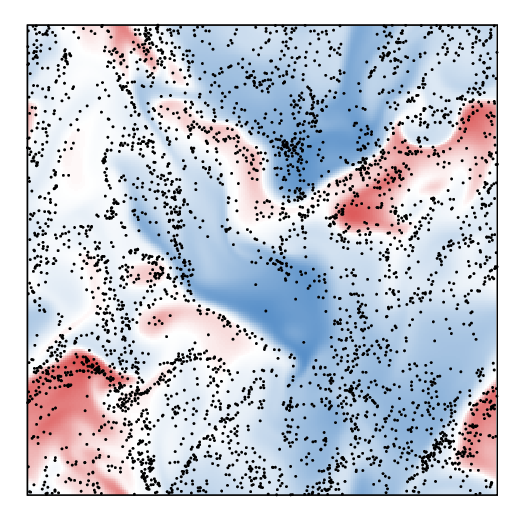}
  \put(10,85){\textbf{(a)}}
  \end{overpic}
  \begin{overpic}[scale=0.6,percent]{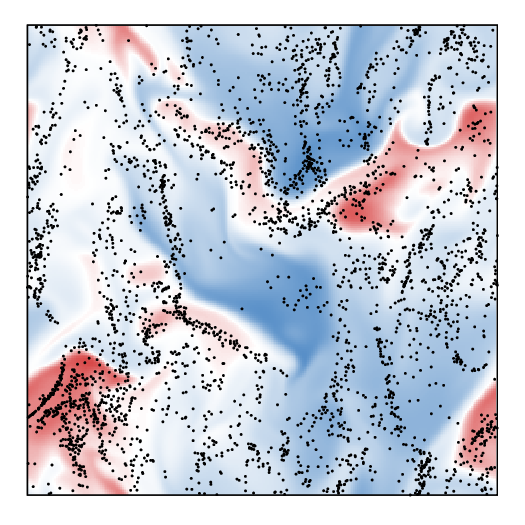}
  \put(10,85){\textbf{(b)}}
  \end{overpic}
  \begin{overpic}[scale=0.6,percent]{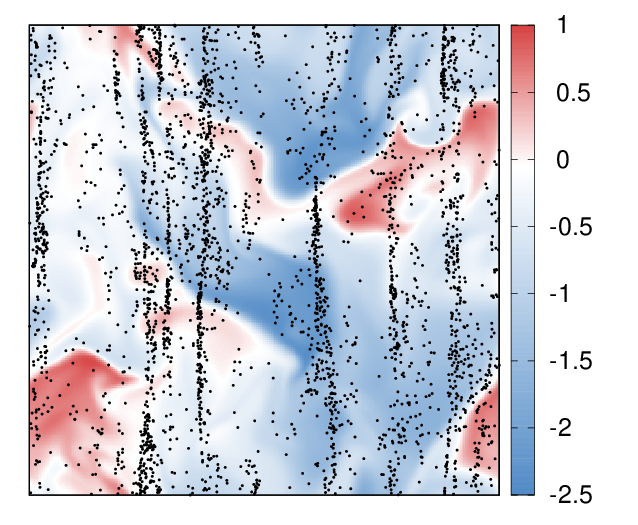}
  \put(9,70){\textbf{(c)}}
  \end{overpic}
 \caption{Cross sections on a $x-z$ plane of particle distributions superposed to the fluid temperature field $\theta/(\gamma L)$, normalized with the mean temperature jump across the periodic box (color scale).  (a) $\mathrm{St}_\eta=0.5$, (b) $\mathrm{St}_\eta=1$ and (c) $\mathrm{St}_\eta=5$. The particles shown are within a distance  $0.01 L$ from the visualized cut on either side in the normal direction. Point sizes are unrelated to the physical particle radius.
 \label{fig:snap}
}
\end{figure}

\subsubsection{Particle settling}
\label{sub.sub.sec.Lag.part.dyn.stat}

We begin our analysis of the dynamics of inertial particles by considering particle velocity statistics as a
function of $St_\eta$. Once the flow properties are fixed, this is the only parameter affecting particle trajectories since, in our simplified model, the temperature of a particle has no effect on its 
density and no feedback on fluid
temperature. The results discussed here are therefore valid for any value of
$\mathrm{St}_\theta$.
In figure~\ref{fig:snap} three vertical cuts of the temperature field are displayed, together with the particle distributions. All three particle distributions are inhomogeneous. Panels (a) and (b) represent intermediate values of $\mathrm{St}_\eta$, where particle inertia is small enough to allow particles to follow fluid structures, albeit with the clear marks of preferential concentration along high-strain regions. Those regions are also characterised by intense temperature fronts. However, particle clusters are shifted downwards with respect to fronts because the finite settling velocity. This effect will be further commented when considering the statistics of temperature along particle trajectories. In panel (c), where $\mathrm{St}_\eta=5$, a different clustering mechanism is active: heavy particles settle so fast as to effectively see the fluid velocity as a time-dependent, two-dimensional field, leading to the formation of vertical streaks \citep{Bec.set.2014}.

Because of gravity, the mean particle vertical velocity is not zero, at variance of the horizontal ones.
In the limit of 
very small and very large inertia $\langle v_{p,3}\rangle$ is dominated by the still-fluid contribution of (\ref{eq.part.mean.vel})
(particles behave either as tracers or move unperturbed through the fluid) so that in both limits 
$\langle u_3 \rangle_p\simeq 0$. At intermediate
$\mathrm{St}_\eta$, preferencial concentration leads to $\langle u_3\rangle_p<0$, which
enhances settling \citep{Bec.set.2014}.

\begin{figure}
  \begin{overpic}[scale=0.6,percent]{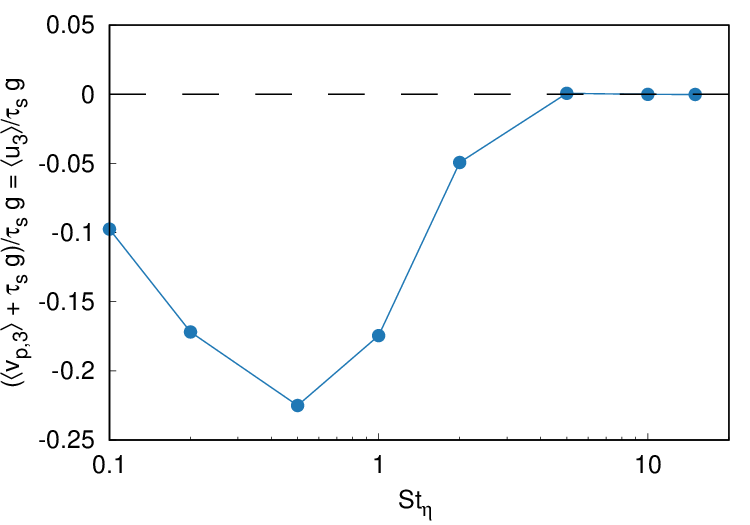}
  \put(80,61){\textbf{(a)}}
  \end{overpic}   
  \begin{overpic}[scale=0.6,percent]{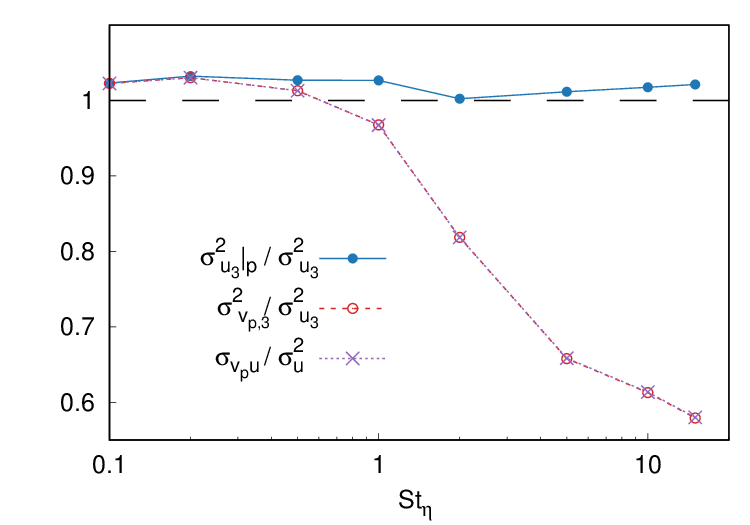}
  \put(80,61){\textbf{(b)}}
  \end{overpic}
 \caption{
 (a) Deviation of the mean value of particle vertical velocities normalized by their respective terminal values in a still-fluid, as a function of $\mathrm{St}_\eta$. (b) Particle and fluid velocity variances and covariances. The variance of particle velocity is completely determined by the covariance of particle and fluid velocity. 
   \label{fig:settling}
 }
\end{figure}

The effects of preferential sampling can be appreciated in
figure~\ref{fig:settling}, where the relative deviation of the average vertical velocity of the particles from the still-fluid terminal speed is
plotted as a function of $\mathrm{St}_\eta$. Such deviation is indeed proportional to the $\langle u_3\rangle_p$ and shows a maximum (in absolute value) around
$\mathrm{St}_\eta\sim0.5$, with particles falling more than $20\%$
faster than in still fluid.  For large inertia, particles fall very fast and correlations with the fluid velocity are lost, so $\langle v_{p,3}\rangle$ tends to be dominated by the still-fluid contribution of eq.(\ref{eq.part.mean.vel}). 
%WHY IS  $\sigma_{v_p}^2>1$ FOR $St_\eta\to 0$? ADD BEHAVIOUR $\sim St$ AND $\sim St^{-2}$ AT SMALL AND LARGE St?

Another important observation comes from \eqref{eq.part.var.vel}, which reveals that the variance of particle velocity is completely determined by the covariance of particle and fluid velocity. This can be seen in figure~\ref{fig:settling}b where the two observables are compared.

\subsubsection{Temperature statistics}
\label{sub.sub.sec.Lag.part.therm.stat}

\begin{figure}
\centering
 \begin{minipage}{0.48\textwidth}
  \begin{overpic}[scale=0.6,percent]{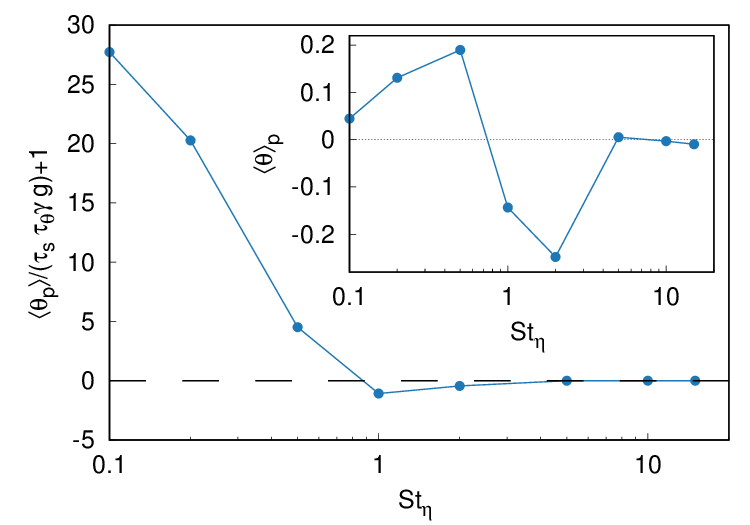}
  \put(80,60){\textbf{(a)}}
  \end{overpic}
\end{minipage}
 \begin{minipage}{0.48\textwidth}
  \begin{overpic}[scale=0.6,percent]{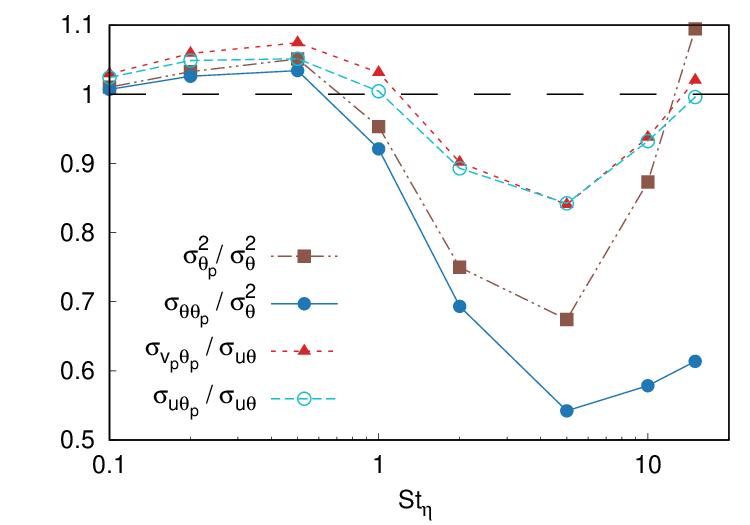}
  \put(80,60){\textbf{(b)}}
  \end{overpic}
 \end{minipage}
 \begin{minipage}{0.48\textwidth}
  \begin{overpic}[scale=0.6,percent]{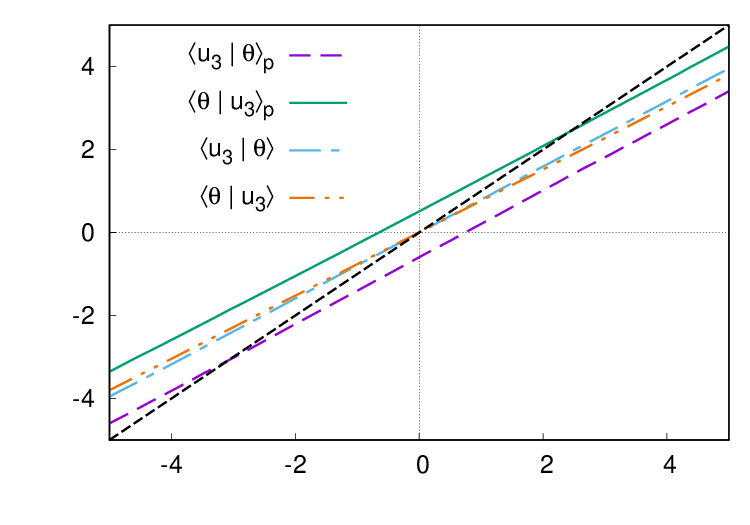}
  \put(80,60){\textbf{(c)}}
   \end{overpic}
 \end{minipage}
 \begin{minipage}{0.48\textwidth}
  \begin{overpic}[scale=0.6,percent]{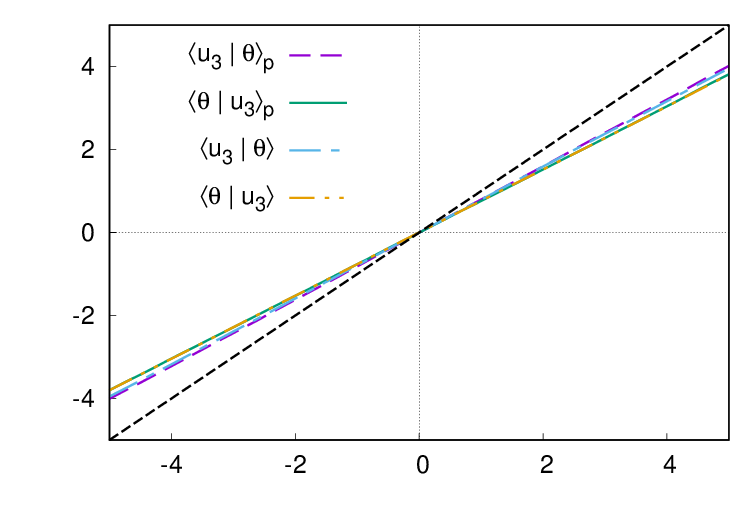}
  \put(80,60){\textbf{(d)}}
  \end{overpic} 
 \end{minipage}
 \caption{Upper panels: (a) Mean values of mean particle temperature normalized by their respective terminal values in a still-fluid, as a function of the dynamic Stokes numbers at fixed ratio $\mathrm{St}_\eta/\mathrm{St}_\theta=1$. Inset: mean fluid temperature as sampled by inertial particles. (b) Particle temperature variance and various covariances. Lower panels: Eulerian and Lagrangian (along inertial particle trajectories) conditional averages of vertical velocity and temperature of the fluid at (c) $\mathrm{St}_\eta=0.5$ and (d) $\mathrm{St}_\eta=5$. 
\label{fig:theta1}
}
\end{figure}

We now look at the statistics of particle temperature. As discussed in section
\ref{sec2}, it is natural to assume that thermal inertia is proportional
to the mechanical one. For this reason in the first part of this analysis we
will vary $\mathrm{St}_\eta$ and $\mathrm{St}_\theta$ by keeping  $\mathrm{St}_\eta/\mathrm{St}_\theta=1$.
Figure~\ref{fig:theta1}a shows the mean relative deviation of particle
temperature with respect to  the terminal temperature $-\tau_s\tau_\theta\gamma
g$. As with the settling velocity, also particle temperature is affected by
preferential sampling, here of fluid temperatures. At large $\mathrm{St}_\eta$
particles trajectories decorrelate from the fluid and particel temperature
converges to the expected, still-fluid value. At intermediate values of inertia,
$\langle\theta\rangle_p\neq 0$ because of preferential sampling. At
$\mathrm{St}_\eta\gtrsim1$ particles preferentially sample downweling regions of the flow, which are also correlated to negative temperature
fluctuations: the resulting particle temperature is lower than the
still-fluid value, in analogy with enhanced settling. 

An unexpected feature which can be appreciated in the inset of
figure~\ref{fig:theta1}a is that the average fluid temperature fluctuation measured along
inertial particle trajectories acquires a positive value for $\mathrm{St}_\eta\lesssim 1$.
This seems at odds with the above mechanism. Figures~\ref{fig:theta1}c and \ref{fig:theta1}d show the average fluid
temperature conditioned on the local value of vertical fluid velocity component
(and {\em vice versa}). The Eulerian statistics is compared with the Lagrangian
one along inertial particle trajectories. Eulerian statistics shows a linear
correlation between vertical velocity and temperature, as expected. This is also true for the Lagrangian statistics, but
for $\mathrm{St}_\eta=0.5$ the latter shows a positive (negative) shift of temperature
(vertical velocity). The shift disappears at larger $\mathrm{St}_\eta$, where inertial
particles decorrelate from the flow, and Lagrangian statistics recovers the
Eulerian one (see figure~\ref{fig:theta1}d).
 This effect can be understood by inspecting figure~\ref{fig:snap}a. In this snapshot, corresponding to the case in figure~\ref{fig:theta1}c, one can see how particles tend to
accumulate at temperature fronts, but overshoot the negative $\theta$ regions and accumulate at the top of underlying, 
positive ones. This phenomenon biases the statistics of the observed
temperatures towards positive $\theta$. This bias disappears at large
$\mathrm{St}_\eta$ (figs.~\ref{fig:snap}b and \ref{fig:snap}c), as particles fall faster and clusters decorrelate from the features of the temperature field.

%==================================

The variance of
temperature fluctuations is considered in figure~\ref{fig:theta1}b. At small-to-moderate
$\mathrm{St}_\eta$ particle temperature fluctuations are increasingly attenuated by the
filtering effect of thermal inertia, similarly to what is observed for
velocity. However, at larger $\mathrm{St}_\eta$, an increase in $\sigma^2_{\theta_p)}$ is
observed. In order to rationalize this phenomenon we analyzed the different
contributions to $\sigma^2_{\theta_p}$.

As can be
seen from eq.~(\ref{eq.part.var.temp}), the variance of particle temperature is
the sum of two contributions. The first is the covariance of particle and fluid
temperature and the second is proportional to the covariance of particle
velocity and temperature. The former shows, after a progressive
decorrelation at low and intermediate $\mathrm{St}_\eta$, an increase at large $\mathrm{St}_\eta$. On the
other hand, the
latter term is further multiplied by $\mathrm{St}_\eta$ and becomes therefore relevant only at
large $\mathrm{St}_\eta$. For $\mathrm{St}_\eta=15$, the maximum value considered here, the two contributions become comparable.

In turn, $\sigma_{v_p\theta_p}$ can be written as the sum of three covariances. If $\tau_s=\tau_\theta$ the relation \eqref{eq.part.var.covar} can be written more transparently as
\begin{align}
\sigma_{v_p \theta_p} &=
\frac{1}{2}  \left [    \sigma_{u \theta_p  } 
+  \sigma_{v_p \theta}
+\tau_s \gamma \sigma_{u v_p } \right ]
\label{eq.var.st1}
\end{align}
It can clearly be seen from figure~\ref{fig:theta1}b that $\sigma_{u\theta_p}\simeq\sigma{v_p\theta_p}$ for most of the range analysed. This implies that also $\sigma{v_p\theta}+\tau_s\gamma\sigma_{uv_p}\simeq\sigma{v_p\theta_p}$ so that essentially all terms contribute in a comparable way.

\begin{figure}
  \begin{overpic}[scale=0.6,percent]{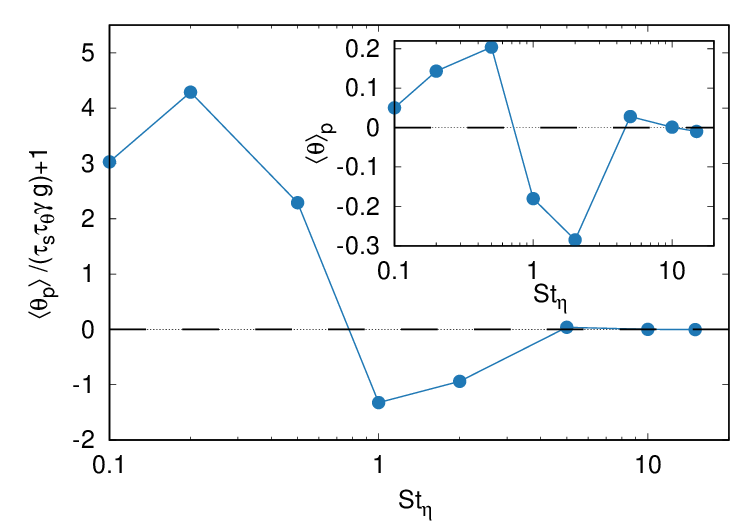}
    \put(17,60){\textbf{(b)}}
  \end{overpic}
  \begin{overpic}[scale=0.6,percent]{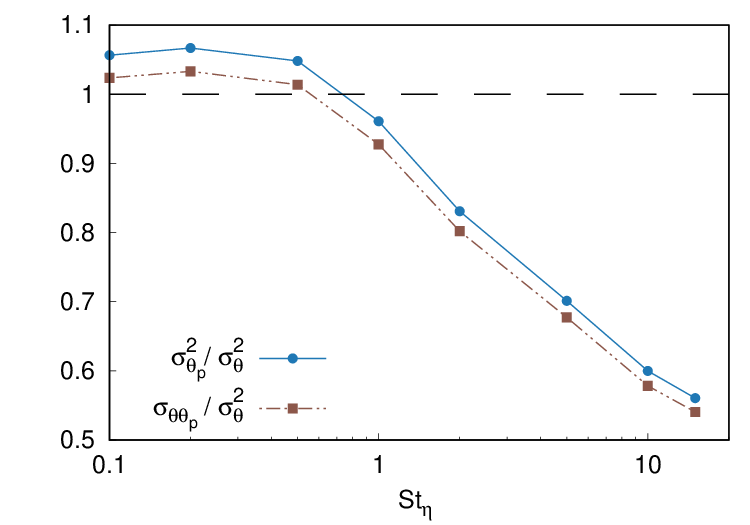}
    \put(82,60){\textbf{(a)}}
  \end{overpic}
 \caption{(a) Mean values of mean particle temperatures normalized by their respective terminal values in a still-fluid, as a function of the dynamic Stokes number $\mathrm{St}_\eta$ at fixed thermal Stokes number $\mathrm{St}_\theta=1$. Inset: mean fluid temperature as sampled by inertial particles. (b) Particle temperature variance and covariance of fluid and particle temperatures. In this case the latter is the dominant term, explaining almost completely the fluctuations in particle temperature. 
 \label{fig:theta2}
}
\end{figure}

In order to discriminate the effects of dynamic and
thermal inertia we performed two sets of simulations by fixing $\rm{St}_\theta$ (respectively $\rm{St}_\eta$)  and letting $\rm{St}_\eta$ (respectively, $\rm{St}_\theta$) vary.
The behaviour of the system with fixed thermal inertia $\mathrm{St}_\theta=1$ at varying
$\mathrm{St}_\eta$ is easily understood. Since we consider only one-way coupling both in
temperature and in momentum, particle dynamics is only affected by 
inertia, so the velocity statistics is the same as in the previous case, as well as
the statistics of the fluid temperatures observed by the particles. However,
since temperature inertia has a filtering effect on the contribution coming from fluid thermal fluctuations, particle temperatures are
affected by this choice. 
The positive values in $\langle\theta_p\rangle$ at small
$\mathrm{St}_\eta$ are still visible in figure~\ref{fig:theta2}a, confirming that
they are due to effects of preferential sampling.
On the other hand, figure~\ref{fig:theta2}b shows that the behaviour of the temperature variance is
qualitatively different from the previous case. When $\mathrm{St}_\theta$ is kept fixed, the main effect of
particle inertia, at large $\mathrm{St}_\eta$, is to decorrelate the Lagrangian
statistics from the Eulerian one. With a fixed thermal inertia, also the other
terms in equations \eqref{eq.part.var.temp} and \eqref{eq.part.var.covar}
decrease, leading to a decreasing $\sigma^2_{\theta_p}$ at large
$\mathrm{St}_\eta$.

We analyzed the same quantities by keeping $\mathrm{St}_\eta$ fixed and changing thermal
inertia. 

\begin{figure}
  \begin{overpic}[scale=0.6,percent]{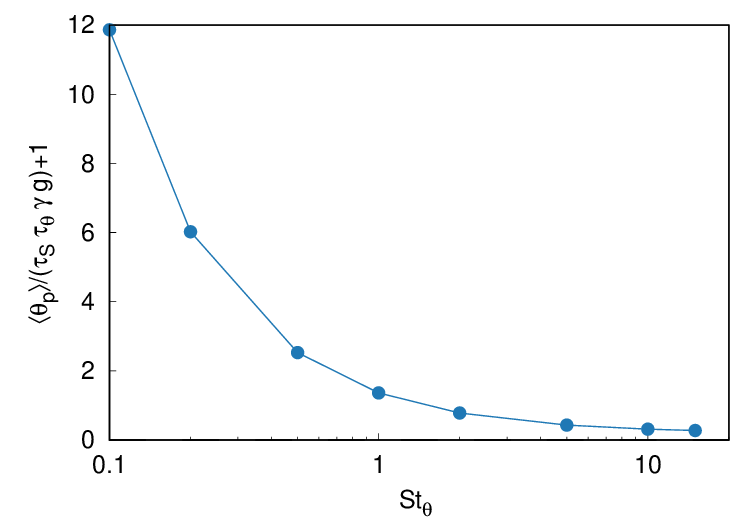}
  \put(80,62){\textbf{(a)}}
  \end{overpic}  
  \begin{overpic}[scale=0.6,percent]{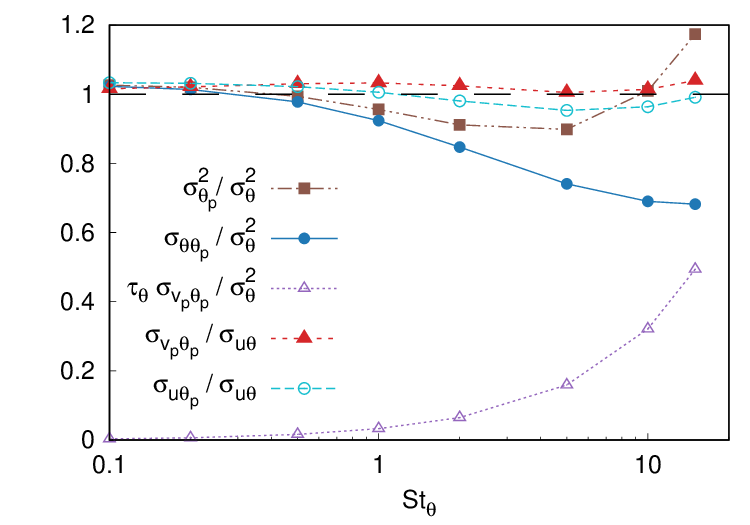}
  \put(80,62){\textbf{(b)}}
  \end{overpic}
 \caption{
 (a) Mean values of mean particle temperatures normalized by their respective terminal values in a still-fluid, as a function of the thermal Stokes number $\mathrm{St}_\theta$ at fixed thermal Stokes number $\mathrm{St}_\eta=1$. (b) Particle temperature variance and various covariances.
 \label{fig:theta3}
}
\end{figure}

Concerning the mean particle temperatures, figure~\ref{fig:theta3}a shows that
particle temperature relaxes monotonically to a value close to the terminal
temperature: this confirms that the non-monotonic behaviour observed in the
other sets of simulations was indeed caused by a non-trivial interplay of
preferential concentration and a finite settling velocity. 

Figure~\ref{fig:theta3}b shows, at large $\mathrm{St}_\theta$, a similar behaviour
to the one observed in the runs with $\mathrm{St}_\theta=\mathrm{St}_\eta$,
with an increase in particle temperature variance, driven by the increase in
$\tau_\theta$. Indeed, the covariance
$\sigma_{\theta\theta_p}$ between particle and fluid temperature decreases at large $\tau_\theta$,
as it should because of the damping due to thermal inertia. The driving
contribution is instead coming from $\gamma\tau_\theta\sigma_{v_p\theta_p}$:
since the correlation between particle temperature and velocity remains almost
constant, the whole term increases about linearly at large $\mathrm{St}_\theta$,
causing the increase in particle temperature fluctuations. Notice also that the
approximate equivalence $\sigma_{v_p\theta_p}\simeq\sigma_{u\theta_p}$ is still
valid, but here (see eq.\eqref{eq.part.var.covar}) the coefficient $\tau_\theta/(\tau_\eta+\tau_\theta)\to 1$ for
$\tau_\theta\gg\tau_\eta$, implying that in this limit the other two terms become subdominant.

\section{Conclusions}
We analysed the statistics of particle velocity and temperature for inertial particles falling in the bulk of a turbulent convective flow. In order to do this we simulated homogeneous convection by imposing a steady vertical temperature gradient across the domain. We confirm that the average settling velocity is affected by preferential sampling of fluid velocity, which in our simulations always favours negatively biased vertical velocities. The clear consequence of this is that, at least for the model studied, turbulent settling is always faster than in a still fluid.

An analysis of covariances allow us to trace the origin of the fluctuations in particle temperature and velocity. Particle velocity variance is found to be identical to the correlation between particle and fluid velocities. The situation is more complex for the statistics of particle temperature where a few unexpected behaviours were found. Indeed, contrary to what preferential sampling would suggest, the average particle temperature was found to be larger than the mean gradient for an intermediate range of $\mathrm{St}_\eta\lesssim{1}$. The reason of this behaviour lies in a non-trivial competition between preferential concentration, preferential sampling and settling. Heavy particles tend to concentrate in regions of high strain, which are also characterised by strong temperature gradients \citep{bec2014clustering}, at the border between hot and cold fluid. On the other hand, preferential sampling favours regions of negative vertical fluid velocity and low temperatures. Settling produces an overshoot of the border between cold and hot regions, and therefore a bias towards relatively hotter fluid. We further analysed how the correlations between velocities and temperature 
%(both the particles' own and the ones of the fluid, as sampled by the particle) 
determine the variance of particle temperature. 

In perspective, it will be interesting to extend our study beyond the one-coupling regime and investigate how particle feedback on the fluid changes the observed effects.

\section*{Acknowledgments}
The authors acknowledge HPC CINECA for computing resources within the INFN-CINECA Grant INF25-fldturb.

\section*{Declaration of Interests}
The authors report no conflict of interest.

%\section*{References}
\bibliographystyle{jfm}
\bibliography{bib_particles_unstable_stratification.bib}

\end{document}